\documentstyle[prl,aps,epsf,multicol]{revtex}
\begin{document}

%%%%%%%%%%%FIGURES%%%%%%%%%%%%%%%%%%
%
\newcommand{\fig}[2]{\epsfxsize=#1\epsfbox{#2}}
% 
%
%%%%%%%%%DEUX COLONNES%%%%%%%%%%%%%%
%   
\newcommand{\passage}{%%
\end{multicols}\widetext\noindent\rule{8.8cm}{.1mm}%
  \rule{.1mm}{.4cm}} 
 \newcommand{\retour}{%%
 %        \hspace{.2cm}
\noindent\rule{9.1cm}{0mm}\rule{.1mm}{.4cm}\rule[.4cm]{8.8cm}{.1mm}%
         \begin{multicols}{2} }
 \newcommand{\unecol}{\end{multicols}}
 \newcommand{\deuxcol}{\begin{multicols}{2}}
%
%
%%%%%%%%%%%%%%%%%%%%%%%%%%%%%%%%%%%%
%
%
\newcommand{\beq}{\begin{equation}}
\newcommand{\eeq}{\end{equation}}
\newcommand{\beqa}{\begin{eqnarray}}
\newcommand{\eeqa}{\end{eqnarray}}
%
%\cst {\rm Cst}

%\setcounter{page}{1}

\tolerance 2000

\author{Pascal Chauve{$^1$}, Pierre Le Doussal{$^2$} and Thierry
Giamarchi {$^1$} 
}
\address{{$^1$} CNRS-Laboratoire de Physique des Solides, Universit{\'e} de
Paris-Sud, B{\^a}t. 510 , 91405 Orsay France}
\address{{$^2$}CNRS-Laboratoire de Physique Th{\'e}orique de 
l'Ecole Normale Sup{\'e}rieure,
24 rue Lhomond,75231 Cedex 05, Paris France.}

\title{Dynamical Transverse Meissner Effect and Transition in Moving Bose Glass}
% \date{\today}
\maketitle

\begin{abstract}
We study moving periodic structures in presence of correlated 
disorder using renormalisation group.
We find that the effect of disorder persists
at all velocities resulting at zero temperature in a Moving Bose Glass 
phase with transverse pinning. At non zero temperature we find 
{\it two} distinct moving glass phases. We predict a sharp transition, as velocity 
increases between the Moving Bose Glass, where the transverse Meissner effect
persists in the direction transverse to motion and a Correlated Moving Glass 
at high velocity, where it disappears. Experimental consequences for vortex
lattices and charge density wave (CDW) systems are discussed.
\end{abstract}

%\pacs{to be added}

%\narrowtext

\deuxcol

The interplay between elastic or plastic
properties of periodic systems and quenched disorder,
relevant to many experimental systems, has been proposed
to lead to several static glass phases with 
complex ground states, pinning barriers and slow 
creep dynamics. Experiments on vortex lattices
\cite{zeldov} support a transition upon increase of point disorder
between a dislocation-free quasi-ordered Bragg glass (BrG)
\cite{tgpldbragg} and an amorphous vortex glass VG \cite{fisher}.
Correlated disorder, e.g. heavy ion columnar tracks, leads to
a stronger glass, termed the Bose glass (BoG) by analogy 
to localized 2D bosons \cite{nv,balents,gtpldcorr}.
The hallmark of the static BoG, compared to an anisotropic VG,
is the transverse Meissner effect (TME) \cite{nv}:
the flux lines being localized along the columns,
the equilibrium response to a perpendicular field vanishes
below a threshold $H^c_\perp(T)$. Although there is experimental evidence
for a liquid to BoG transition at $T_{BoG}$
with anomalous angular dependences of $T_{BoG}$ and transport
\cite{delacruz},
only recently have attempts been made 
\cite{smith_tme} to observe the TME {\it directly}.

The dynamical states for {\it moving} periodic structures 
in presence of point disorder
have been investigated recently 
\cite{koshelev_vinokur,tgpldmgprl,tgpldlong,balents_marchetti}.
The naive expectation is that a 
fast moving system averages out disorder,
resulting only in an increase of the effective temperature.
It was found instead \cite{tgpldmgprl,tgpldlong}
that the residual periodicity
transverse to motion still sees static disorder, and glassy features
survive, such as {\it transverse pinning},
leading to a moving glass (MG) state. The system was shown to
be well described by the MG model
\cite{tgpldmgprl,tgpldlong} which
involves only transverse displacements.
Rows of a vortex lattice driven by a Lorentz force should thus flow 
along well defined static channels, with barriers to transverse motion
and a transverse depinning threshold $F_c^t(v)$
\cite{tgpldmgprl}.
This was confirmed in numerical simulations \cite{moon,olson}.
Bitter decoration experiments 
and STM imaging \cite{kes_nature} show
vortex lattices moving along their principal axis
and forming stable channel structures
either fully coupled (no dislocations) at large velocity $v$ or
decoupled (with smectic order) at lower $v$ \cite{kes_nature}, as
found in simulations \cite{olson}. Transverse barriers could
also explain the anomalously small Hall effect in Wigner solids
\cite{williams_transverse}. Finally, CDW systems
in a steady current were observed to exhibit a depinning
threshold under an electric field applied along their periodicity direction
transverse to carrier motion \cite{vanderzant}. It provides
a direct measurement of the transverse critical force $F_c^t(v)$,
observed \cite{vanderzant} to decay exponentially
with $v$ in $d=3$, as predicted in \cite{tgpldmgprl}.

These studies having focused on point disorder
it is natural to ask whether lattices moving
in presence of {\it correlated} disorder also retain 
some glassy features of the static BoG.
One question is whether the transverse periodicity 
results in a persistence of the TME when the field is
applied transverse to motion. Additional physics compared to 
point disorder is expected since the localization effect 
of the columns (reducing thermal
wandering) compete with heating by motion. Probing these effects
as well as vortex correlations is experimentally accessible.

In this Letter we investigate moving lattices in presence of 
correlated disorder. We describe the system within the MG
model which involves only transverse displacements.
Using renormalization group (RG) 
we find that the effect of weak disorder persists at all $v$
resulting at $T=0$ in a moving Bose glass 
phase (MBoG) with transverse pinning and TME.  At $T>0$ 
we find {\it two} distinct moving glass phases.
Our model exhibits a sharp transition,
as velocity increases, between the MBoG 
where the TME persists and a higher velocity phase,
the Correlated Moving Glass (CMG), where it disappears.
While in the CMG the {\it thermal} fluctuations 
grow logarithmically with distance, in the MBoG 
they remain bounded as the channels remain localized.
The properties of these moving phases and
transition are summarized in Table 1 and should be
testable in vortex and CDW systems \cite{footnotecdw}.

\begin{figure}
\begin{tabular}{c|c|c|c}
\hline
phase & Response & $u_{stat}$ & $u_{th}$  \\
\hline
MBoG & & & \\
$T=0$ $v>0$ or& TME & $x^{1/4}$ & bounded \\
$T>0$ $v<v_{c}(T)$ & & $y^{1/2}$ & $ u_{th} \to l_\perp $ \\
\hline
transition $v_c^-$ & $ H^y_c \rightarrow $ cst &&
$l_\perp \rightarrow $ cst \\
\hline 
\hline
transition $v_c^+$ & $c_{44}\sim \eta \sim e^{C/
(v-v_{c})^{\alpha}}$ && delocalized \\
\hline
CMG & & $x^{1/4}$ & logarithmic \\
$T>0$, $v>v_{c} (T)$ & no TME & $y^{1/2}$ & $ l_\perp=+\infty $\\
\hline
\end{tabular}
\label{table}
\vspace{1mm}\noindent
\small {\bf Table I}:
{Characteristics of the two moving phases}

\end{figure}

Since heavy ions tracks act as strong pinning centers, except at higher $T$
or fields, it is also important here, as in the case of point disorder,
to investigate the effect of dislocations. At large enough $v$ the effect 
of disorder is strongly reduced and the 
system should recover a large degree of topological order.
Point disorder studies in 2D \cite{moon,olson} suggest by analogy
to straight lines in columnar disorder that
transverse periodicity (even in the presence of few dislocations)
survives down to relatively low velocity.
Hence, we expect the transverse MG model based
on the channel structure to be good starting point.

We consider a lattice moving over a substrate
with weak gaussian disorder correlated along $z$.
The velocity $v$ is along a principal lattice 
direction $x$. The MG model only assumes topological order in 
the direction transverse to motion $y$ and consists in the equation 
of motion for the component of the displacement field along $y$,
denoted $u_{rt}$, $r=(x,y,z)$.
It describes both (i) fully elastic flow, coupled channels 
(weak disorder or large $v$) (ii) flow with phase slips (dislocations)
occuring between channels (stronger disorder or intermediate $v$)
and reads:
\begin{eqnarray}
(\eta \partial_t + v \partial_x - c_x \partial_x^2
- c_y \nabla_y^2 - c_z \partial_z^2) u_{rt} \nonumber \\
= F[x,y,u_{rt}] + \zeta_{rt} 
\label{eqmo}
\end{eqnarray}
with $\langle \zeta _{rt} \zeta _{r't'} \rangle = 2 \eta T \delta^d(r-r') \delta(t-t')$ and the correlator of the static pinning force along $y$ is
$\overline{F[x,y,u] F[x',y',u']} = \delta(x-x') \delta^{d_y}(y-y') \Delta(u-u')$.
The bare $\Delta(u)$, defined in \cite{tgpldlong},
is of range $r_f$ and has the lattice period $a$.
We denote the spatial dimension of $y$ by $d_y$,
the case of physical interest being $d=2+d_y=3$.
The bare friction $\eta_0$ is absorbed in $v$.
The $c_i$ can be estimated for a flux lattice
with the field along $z$ \cite{footnote8}.
Adding a transverse field $H^y$ along $y$ amounts to add in (\ref{eqmo})
a surface force $h^y(\delta(z-L_z/2)-\delta(z+L_z/2))$, $h^y \propto H^y$.

Let us first analyze the effect of disorder using first order 
perturbation theory. Correlations split into static disorder-induced 
displacements and thermal displacements $\overline{\langle |u_{q,\omega}|^2 \rangle} = 
\delta(\omega) \delta(q_z) C_{stat}(q_\perp) + C_{th}(q,\omega)$.
The static displacements are $z$ independent and identical to the case of 
point disorder in $d-1$ dimensions:
\begin{eqnarray}
\overline{\langle (u_{r t}-u_{0 t'})^2 \rangle}\simeq \Delta(0)
\int_{q_x,q_y} \frac{2(1-\cos(q_x x + q_y y))}{
(c_x q_x^2 + c_y q_y^2)^2 + v^2 q_x^2}
\label{static}
\end{eqnarray}
They become unbounded for $d < d_{uc}=4$ ($d_{uc}=3$ for point disorder).
As in \cite{tgpldmgprl} simple perturbation theory breaks 
down beyond a {\it dynamical
Larkin length} $R^y_c(v)$ (at which $u_{stat} \sim r_f$)
which interpolates between the static Larkin length
$R^y_c(v=0) \sim (1/\Delta(0))^{1/(5-d)}$ and the large $v$
estimate (from (\ref{static})) $R^y_c(v) \sim (c_y v r_f^2/
\Delta(0))^{1/(4-d)}$.

A characteristic feature of the static BoG is the {\it reduction}
of {\it thermal} displacements by disorder (while they are unaffected
by point disorder) resulting in the upward shift of the melting line \cite{nv}.
To analyze the competition between these localization effects and
heating by motion we compute equal time thermal 
displacements at low $T$:
\begin{eqnarray}
C_{th}(q,t=0) = (1 + |\Delta''(0)| G_v(q)) T/(c_i q_i^2)
\end{eqnarray}
We find \cite{us_long} that $G_v(q)$ {\it changes sign}
as a function of $v$: it is negative
(reduced thermal displacements) for $v<v^*(q)$ and positive 
(heating by motion) when $v >v^*(q)$.
Setting $1/q$ to be the Larkin length raises the 
possibility of a dynamical transition at $v_c$ where
heating by motion \cite{footnote3} wins over the 
localization by the columns. 

To describe the system beyond the Larkin length
we now use the RG on the dynamical
field theory associated to (\ref{eqmo}). Reducing the 
cutoff $\Lambda_l = \Lambda e^{-l}$ for $q_y$ we get \cite{us_long}:
\begin{eqnarray}
\partial_l \ln c_z &=& \partial_l \ln \eta
= - \frac{\tilde{\Delta}''(0)}{1+\tilde{v}^{2}e^{2l}} \label{eqcz} \\
\partial_l \ln \tilde{T} &=& -d_y + \tilde{\Delta}''(0)
\frac{\frac{1}{2}-\tilde{v}^{2}e^{2l}}{1+\tilde{v}^{2}e^{2l}} \label{eqt} \\
 \partial_l \tilde{\Delta}(u) &=& (2-d_y
+ \frac{1}{1+\tilde{v}^{2}e^{2l}} ) \tilde{\Delta}(u) +
\tilde{T} \tilde{\Delta}''(u) \nonumber \\
&& + \tilde{\Delta}''(u) 
(\tilde{\Delta}(0) - \tilde{\Delta}(u))
-\frac{\tilde{\Delta}'(u)^2}{1+\tilde{v}^{2}e^{2l}} \label{eqdel}
\end{eqnarray}
$c_x$, $c_y$, $v$ are not renormalized,
$\tilde{v}=v/(2 \Lambda \sqrt{c_{x}c_{y}})$,
$\tilde{T}_l = T_l C_l \Lambda_l^{d_y}/\sqrt{c_x c_z(l)}$
the reduced temperature \cite{footnote4} and 
$\tilde{\Delta}_{l}(u)=S'_{d_y} 
\Lambda_l^{d_y-3} \Delta _{l}(u)/4 c_{y}\sqrt{(1 + \tilde{v}^{2}e^{2l}) c_{x} c_{y}}$.
These equations reveal three phases and a transition as follows:

{\it Static Bose Glass}: at $v=0$ (\ref{eqcz},\ref{eqt},\ref{eqdel})
describe dynamically an elastic version of the BoG similar 
to the one studied at equilibrium in \cite{balents,gtpldcorr}.
The model exhibits analytically many 
of the properties of the BoG induced by strong columnar defects
e.g. the TME \cite{balents}. At $T=0$ the ground state is $z$-independent,
thus the problem is identical to point disorder in $x,y$ space
with identical RG equations. Beyond the Larkin length 
$R_c \sim a e^{l_c}$, $\tilde{\Delta}_{l}(u)$ develops a cusp
$\tilde{\Delta}''_{l}(0) \to -\infty$ giving rise to a depinning threshold.
From (\ref{eqcz}) the tilt modulus $c_z$
diverges at $R_c$ implying a vanishing linear response to
$H_\perp$ and leading to the TME \cite{balents}. The 
$T=0$ fixed point reads \cite{tgpldbragg,balents}, to ${\cal O}(\epsilon=5-d)$,
$\tilde{\Delta}_{BoG}^*(u)=\frac{\epsilon}{6} (\frac{a^2}{6}-u(a-u))$ (for $0<u<a$).
At $T>0$ we observe \cite{footnote5} from our RG equations 
a remarkable feature compared to point disorder: 
the effective temperature $\tilde{T}_l$ 
runs to $0$ at a {\it finite} length scale $R_{loc} > R_c$.
This is consistent with localization effects, expected in the BoG,
setting in beyond $R_{loc}$.

{\it Moving Bose Glass}: in the moving system, at $T=0$ for any $v$
{\it and} at $T>0$ for $v<v_c(T)$ defined below, the RG flows to a 
$T=0$ fixed point, which we call the MBoG. The RG equation 
(\ref{eqdel}) at $\tilde{T}_l=0$ resembles the
one for the MG with point disorder in $d-1$ dimensions \cite{tgpldlong}.
To ${\cal O} (\epsilon =4-d)$ the asymptotic behavior for any initial velocity
is $\tilde{\Delta}_l(0)-\tilde{\Delta}_l(u) \to \frac{\epsilon}{2} u(a-u)$ 
while $\tilde{\Delta}_l(0) \sim e^{\epsilon l}$. Since
$\tilde{\Delta}_l(u)$ develops a cusp at the {\it dynamical Larkin length} 
$R^y_c(v)$ \cite{footnote6} the MBoG is also characterized by
transverse barriers and a $T=0$ transverse depinning threshold 
$F^t_c(v) \sim c_y r_f/R_c^y(v)^2$: it can be checked by adding
a small transverse force and is manifest from the divergence of
the relaxation time $\eta_l$ at $R^y_c(v)$ (from \ref{eqcz}). The novel feature,
specific to correlated disorder, is that the linear response to 
a transverse field $H^y$ along $y$ vanishes as the tilt modulus $c_{z}(l)$ 
diverges beyond $R^y_c(v)$ (from (\ref{eqcz})).
Since a finite $H^y$ acts as an additional surface force $h^y$,
the existence of a transverse depinning threshold $F^t_c(v)>0$
implies that the field cannot penetrate from the surface inside the bulk.
Thus the TME persists in the MBoG, the $T=0$ threshold field 
$H_c^y$ being related to $F_c^t$ through $h_c^y(v) \sim R_c^z F_c^t 
\sim \sqrt{c_y c_z} r_f/R_c^y(v)$ where the penetration
length along $z$ is $R_c^z(v) \sim \sqrt{c_z/c_y} R_c^y(v)$.

Positional correlations in the MBoG are found to be dominated 
by a {\it correlated random force} which yields static,
$z$-independent displacements growing as $u \sim y^{1/2} \sim x^{1/4}$,
i.e as in the point disorder MG at $T=0$ in $d=2$ (and faster 
than in the elastic BoG where $u \sim \ln y$).

\begin{figure}[htb]
\centerline{ \fig{6cm}{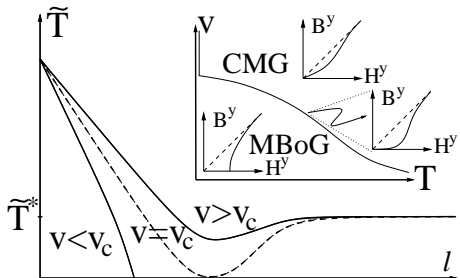} }
\caption{
{\narrowtext  Flow of the effective temperature $\tilde{T}_l$.
Dotted line: separatrix between the MBoG $v<v_{c}$ and the CMG $v>v_{c}$.
Inset: responses and $v,T$ phase diagram of the MG model.}}
\label{fig1}
\end{figure}

We now study the stability of this phase to temperature.
As long as $\tilde{T}_l >0$, 
$\tilde{\Delta}_l(u)$ is analytic ($\tilde{\Delta}''_l(0)<0$).
Remarkably, the coefficient 
of $\tilde{\Delta}''_l(0)$ in (\ref{eqt}) changes sign 
as a function of $v$. Thus $\tilde{T}_l$ decreases at small
$v$ and increases at large $v$, consistent with
the above first order estimate. Our RG shows that 
this competition between localization and heating 
leads to a sharp transition in (\ref{eqmo}) at $v=v_c(T)$. Indeed,
as depicted in fig. \ref{fig1}, for
$v < v_c$, $\tilde{T}_l$ decreases to $\tilde{T}_l=0$ at a finite scale 
$R_{loc}(T,v)= a e^{l_{loc}}$ and remains exactly zero thereafter.
Thus all properties beyond $R_{loc}$ are governed 
by the $T=0$ MBoG fixed point. For $v > v_c$ heating wins
($\tilde{T}_l$ never vanishes), driving the system to
the CMG phase described below. 

Thus the MBoG $T=0$ fixed point is {\it stable} 
to temperature for $v < v_c$. This is specific to correlated disorder and
is in contrast with the MG with point disorder. We surmise that, as in the static BoG,
barriers also diverge in the MBoG, leading to transverse creep,
which would be interesting to check numerically. Localization effects
survive in the MBoG as seen from equal time connected fluctuations 
along $z$, $l_\perp^2 = \lim_{|z-z'| \to \infty}
\frac{1}{2}\overline{\langle (u_{x y z t} - u_{x y z' t})^2 \rangle_{c} }
= \int_{l} \tilde{T}_l$,
which are bounded since the integral only runs up to $l_{loc}$. 
To confirm the persistence of Bose Glass features in
moving systems we also studied numerically the $d_y=0$ version 
of (\ref{eqmo}) (known to exhibit BoG and TME at $v=0$ 
in equilibrium \cite{toy}). Results
are consistent with a TME at $v,T>0$.

{\it Correlated Moving Glass}: as shown in fig. \ref{fig1}
for $T>0$, $v>v_{c}(T)$, the RG flows to a finite temperature fixed point
(perturbative in ${\cal O}(\epsilon =4-d)$)
corresponding to a novel dynamical phase, the CMG.
It is characterized by $\tilde{T}^{*} \sim \frac{\epsilon^2 a^{2}}{16 |\ln \epsilon|}$
and by an analytic fixed point \cite{us_long} for
$\tilde{\Delta}_{l} (u)-\tilde{\Delta}_{l} (0)$.
A correlated random force $\tilde{\Delta}_{l} (0)\sim
e^{\epsilon l}$ is also generated. This behavior is analogous, 
but not identical to the MG fixed point \cite{tgpldlong}
with point disorder, marginal in $d=3$ while the CMG is well below
its upper critical dimension ($\epsilon =1$). Contrarily to the BoG and 
the MBoG, no cusp singularity occurs.
From (\ref{eqcz}) one hence finds that the system responds linearly 
to transverse forces and tilting fields, with
{\it finite} renormalized coefficients $\eta^{R}$ and $c_{z}^{R}$.
However, since at $T=0$ the system instead flows to the MBoG fixed point,
one does expect strongly non linear transverse $I-V$ and 
$B^y-H^y$ characteristics and diverging $\eta^{R}(T),c_{z}^{R}(T)$
as $T \to 0$. Due to the correlated random force, the static roughness 
of the channels is $u \sim y^{1/2}\sim x^{1/4}$ as in the MBoG
(with a smaller amplitude). Thermal displacements however, while
bounded in the MBoG, grow logarithmically in the CMG, 
as $\overline{\langle (u(y) - u(0))^2 \rangle_c} 
\approx 2 \tilde{T}^* \ln y$. Hence the channels are thermally broadened,
with slow decay of 
$\overline{ \langle  e^{i 2 \pi u_{r,t}/a} e^{- i 2 \pi  u_{0,t}/a} \rangle_c } \sim
y^{-\eta}$, $\eta \approx 0.48$ in $d=3$, in contrast
with the finite-width channels of the MG \cite{footnote7}.

\begin{figure}[htb]
\centerline{ \fig{8cm}{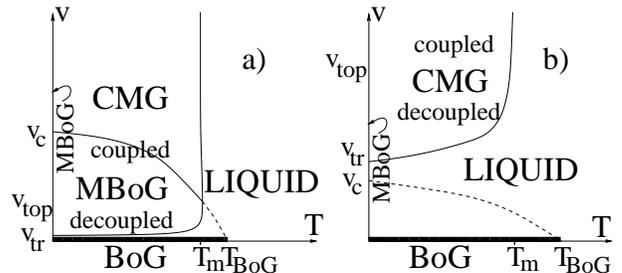} }
\caption{
{\narrowtext Schematic phase diagrams for (a) weak disorder
(b) strong disorder, in $v,T$ plane.}}
\label{fig3}
\end{figure}

{\it Dynamical transition}: although for $T=0$ the moving system is
always in the MBoG phase, for $T>0$ a sharp transition occurs at a 
critical velocity $v_{c} (T)$ between the two above phases.
The separatrix in the RG flow (fig. \ref{fig1})
is parabolic \cite{footnote9}.
Thus the transition occurs when the MBoG length $R_{loc} (T,v)$
equals $R_{0}(v) \equiv \sqrt{2 c_{x}c_{y}}/v$, 
the (here $T$-independent)
length in the CMG at which the flow of $\tilde{T}_l$ reverts.
At low $T$, $R_{loc} (T,v) \approx R_{loc} (0^+,v)=R_{c}^{y}(v)$
and thus \cite{footnote6}
$\eta_0 v_c \simeq 1.3 \sqrt{c_{x} c_{y}}/R_c$
for weak disorder whereas it saturates at 
$\eta_0 v_c \sim \sqrt{c_{x} c_{y}}/a$ for stronger disorder.
For lattices with $c_{66} \ll c_{11}$, \cite{footnote8} yields
$c_y \sim c_{66}$ near $v_c$ and
$\eta_0 v_c \sim c_{66}/\max(R_c,a)$. For weak disorder ($R_c>a$,
$F_c=c_{66} r_f/R_c^2$), one gets $\eta_0 v_c/F_c \sim R_c/r_f$.
(\ref{eqcz}) and the parabolic shape of the separatrix
leads in the CMG to renormalized tilt modulus and friction
diverging as $c_{z}^R \sim \eta^R \sim e^{\frac{{\rm Cst}}{(v-v_c)^\alpha}}$,
$\alpha=\frac{1}{2}$ for $v \to v_c^+$.
It appears as an ordering transition along $z$
with a characteristic length $R^y$ along $y$ remaining finite
while the length $R^z \sim\sqrt{c^R_z/c_y} R^y$ diverges 
as $R^z \sim e^{\frac{{\rm Cst}}{(v - v_c)^\alpha}}$.
However, coming from the MBoG, $l_\perp$ jumps discontinuously
to $l_\perp = +\infty$ in the CMG, and thus the transition
is of mixed discontinuous-continuous character.

%\begin{figure}[htb]
%\centerline{ \fig{6cm}{thetah.eps} }
%\caption{
%{\narrowtext  Expected transverse magnetization versus transverse 
%field characteristics.}}
%\label{fig2}
%\end{figure}

Since the MG model relies on the transverse order
it describes the system for $v > v_{tr}$ 
at which channels appear. For $v > v_{top}$ these
channels recouple and the lattice recovers a good amount of 
topological order \cite{footnotes}. Thus the 
dynamical transition should be observable 
(fig. \ref{fig3} a) for $v_c > v_{tr}$ (weaker disorder) while
it is rounded by plastic effects for $v_c < v_{tr}$
(stronger disorder). Lindemann estimates of $v_{top}$ can be obtained 
\cite{us_long} as in \cite{tgpldlong}
(for $c_{66} \ll c_{11}$, $v_{tr} \sim v_{top}$).
We consider columnar defects of spacing $d$, of 
strength $u_0 \equiv U_0(T)/\epsilon_0$, with
$r_f=l_\perp(T)$($=b_0$ at low $T$), using \cite{nv}.
When the decay length of translational order $R_a$
in the static BoG is such that $R_a > a$, i.e
$d/a > u_0 (r_f/a)^{3/2}$, the MBoG and the 
transition at $\eta_0 v_c \sim \epsilon_0/a^3$
should be observable. For stronger disorder the CMG
exists (fig. \ref{fig3} b) for $\eta_0 v > \eta_0 v_{tr}
\sim U_0\sqrt{r_f/a}/(a^2 d)$. These should be compared to
$F_c \sim U_0/(r_f a^2)$, the single vortex pinning longitudinal threshold.

The plastic limit at stronger disorder can be 
investigated extending the creep arguments of \cite{nv}.
We find that the penetration of $H^y$ results from 
(super)kinks nucleated at the $\pm$ surfaces with a bias 
along $\pm y$ and propagating inside the bulk.
For not very thick samples they produce a small $B^y$
of order the creep velocity. In thick samples
the response is determined by their
competition with (super)kinks nucleated in the bulk
{\it unaffected} by $H^y$, which propagate to the surface.
At $T=0, f>f_c$ the persistence of anomalous tilt 
response depends on whether trajectories of a point in 2D attract
on average \cite{footnote10}.
Thus the absence of $H^y$ penetration in the MG model studied here
results from interactions responsible for the transversely ordered
channel structure. Surface kinks along $y$ will also be nucleated at $T>0$
but in the MBoG they should not penetrate deep in the bulk. 

Additional weak point disorder (known to destabilize the $d=3$ plastic BoG
\cite{toy}) transforms the CMG (above a large length) into a new fixed point 
\cite{us_long} with mixed features: the static roughness of the channels
along $x,y$ grows as in the CMG but is logarithmic along $z$ (delocalization)
with finite thermal width. There is a $T=0$ transverse 
critical force but no TME. At smaller $v$ the MBoG exhibits a greater 
stability to point disorder.

To conclude, some effects of 
correlated disorder on moving periodic structures
(e.g. the existence of two distinct moving phases)
were found to differ radically from those 
due to point disorder. In particular we found a
$T>0$ moving phase (MBoG) with 
strong glassy features.

%\vspace{20cm}
\unecol
%\end{references}


\begin{thebibliography}{999}
\bibitem{footnotecdw} via the mapping
$J_x \sim v$, $E_y \sim f_y$, Bose glass effects, TME and transverse
depinning should also be observable in CDW with correlated disorder
using surface fields.
\bibitem{footnote4}
$C_l=\frac{S'_{d_y}}{2 \pi} 
\rm{asinh} \sqrt{\frac{c_z(l) \Omega^2}{c_y \Lambda_l^2}}$
depends weakly on $l$, $\Omega$ is the cutoff along $z$ and 
$S'_{d_y}$ is the unit sphere area over $(2 \pi)^{d_y}$.
\bibitem{footnote5} As in \cite{chauve_creep} 
$\tilde{\Delta}_{l}(0)-\tilde{\Delta}_{l}(u)=
\tilde{T}_l g(u \chi/\tilde{T}_l)$ satisfies (\ref{eqdel})
with $(1+g)^{2 \prime \prime}+ g - xg' =2$ which has a well behaved solution.
Thus $\tilde{\Delta}''_{l}(0)$ diverges as $-\chi^2/\tilde{T}_l$
and (\ref{eqt}) yields the result.
\bibitem{footnote9}
Using (\ref{eqt}) and a numerical study of (\ref{eqt}, \ref{eqdel}) 
which shows that $- \tilde{T}_l \tilde{\Delta}_l''(0)$ is not singular 
near the transition.
\bibitem{footnote10} while true for an egg carton potential
$V=f(u_x) + g(u_y)$ it is non trivial in general.
\end{thebibliography}
\end{document}